\documentclass[runningheads]{llncs}
\usepackage[T1]{fontenc}
\usepackage{xcolor}
\usepackage{epsfig}
\usepackage{graphicx}
\usepackage[normalsize,it,hang]{subfigure}
\usepackage{amsfonts}
\usepackage{amssymb,hhline}
\usepackage{graphics}
\usepackage{latexsym}
\usepackage{makeidx}
\usepackage{amsmath}
\usepackage{wasysym}
\usepackage{amscd}
\usepackage{color}
\usepackage{rotating}
\usepackage{array}

\begin{document}
\title{
{\small {\tt 23rd International Conference on \\ Computational Methods in Systems Biology (CMSB 2025) \\ 10-12 september 2025, Lyon, France}} \\
\vspace{1cm}A model-free control strategy for selective disruption of parkinsonian brain oscillations}

\titlerunning{Disruption of parkinsonian brain oscillations}
\author{
C\'{e}dric Join\inst{1,6}
\and
Jakub Orłowski\inst{2}
\and Antoine Chaillet\inst{3}
\and
Madeleine Lowery\inst{2}
\and
Hugues Mounier\inst{3}
\and 
Michel Fliess\inst{4,5,6}
}
\authorrunning{C. Join et al.}
\institute{Centre de Recherche en Automatique de Nancy, CNRS, Universit\'{e} de Lorraine, 54506 Vand{\oe}uvre-l\`{e}-Nancy, France \\
{\tt cedric.join@univ-lorraine.fr} \\
\and
School of Electrical and Electronic Engineering, University College Dublin, \\ Dublin, Ireland\\
{\tt \{jakub.orlowski, madeleine.lowery\}@ucd.ie}
\and
Universit\'e Paris-Saclay, CNRS, CentraleSup\'elec, Laboratoire des Signaux et Syst\`emes, 91190 Gif-sur-Yvette, France \\
{\tt \{antoine.chaillet, hugues.mounier\}@centralesupelec.fr}
\and
Laboratoire d'Informatique de l'\'{E}cole Polytechnique, CNRS, \\ 91120 Palaiseau, France \\
{\tt michel.fliess@polytechnique.edu}\\ 
\and
Laboratoire Jacques-Louis Lions, CNRS, Sorbonne Universit\'e, \\ 75005 Paris, France \\
{\tt michel.fliess@sorbonne-universite.fr}\\
\and
AL.I.E.N., 7 rue Maurice Barr\`{e}s, 54330 V\'{e}zelise, France\\
{\{\tt cedric.join, michel.fliess\}@alien-sas.com}
}
\maketitle          
\newpage
\begin{abstract}
Deep brain stimulation (DBS) is an advanced surgical treatment for the symptoms of Parkinson's disease (PD), involving electrical stimulation of neurons within the basal ganglia region of the brain.
DBS is traditionally delivered in an open-loop manner using fixed stimulation parameters, which may lead to suboptimal results.
In an effort to overcome these limitations, closed loop DBS, using pathological subthalamic beta (13--30 Hz) activity as a feedback signal, offers the potential to adapt DBS automatically in response to changes in patient symptoms and side effects.
However, clinically implemented closed-loop techniques have been limited to date to simple control algorithms, due to the inherent uncertainties in the dynamics involved. 
Model-free control, which has already seen successful applications in the field of bioengineering, offers a way to avoid this limitation and provides an alternative method to apply modern control approach to selective suppression of pathological oscillations.

In this paper, we use a computational mean-field model of parkinsonian brain activity to show that model-free control provides selective disruption of pathological beta activity within the network.
We show that this technique successfully detects and suppresses the beta activity, while preserving the non-pathological activity in the gamma ($\geq 30$ Hz) frequency band, even in the presence of extraneous noise.
These results demonstrate the potential for MFC as a viable candidate for closed-loop DBS algorithm in the treatment of Parkinson's disease. 

\end{abstract}

\newpage

\section{Introduction}

Deep brain stimulation (DBS) is an advanced surgical neuromodulation therapy which uses electrodes implanted in the deep structures of the brain to induce therapeutic effects via electrical stimulation.
It is a clinically approved treatment for essential tremor, Parkinson's disease (PD), dystonia and refractory epilepsy, and is approved for obsessive-compulsive disorder as a humanitarian device exemption.  DBS is also being investigated as a potential treatment for a range of conditions including Alzheimer's disease, chronic pain, Tourette's syndrome, major depression, tinnitus and substance abuse.
Since the 2002 FDA approval of DBS for Parkinson's disease, more than 160,000 patients have been implanted with DBS devices worldwide, with numbers continually increasing  \cite{lozano2019}.

Traditionally, DBS in clinical applications has been used in an open-loop fashion, with stimulation parameters (such as pulse amplitude, frequency and  duration) set by a trained clinician in the postoperative period and then updated during follow-up visits.
The open-loop nature of stimulation can lead to suboptimal control of disease symptoms and stimulation-induced side effects.
Overstimulation (prioritizing symptom suppression) can also contribute to reduced battery life, necessitating more frequent replacement operations or more frequent need to recharge the stimulator in case of rechargeable devices.
Despite these limitations, DBS for Parkinson's disease is an  effective theraputic intervention allowing the patients to lower the dosage of the antiparkinsonian medicine and significantly improve their quality of life.

The limitations of open-loop neurostimulation have sparked interest in closed-loop stimulation, which adapts the stimulation parameters based on measurable biomarkers.
Since the first animal \cite{rosin2011} and human \cite{little2013} studies in the early 2010s, closed-loop DBS has consistently been shown to be safe and well-tolerated, providing symptom suppression comparable to open-loop DBS, while reducing incidence of side effects and lowering the stimulation power.
Recent clinical trials with patients undergoing closed-loop DBS for extended periods of time \cite{stanslaski2024} led to the approval for clinical applications of closed-loop DBS for Parkinson's disease in the European Union, the UK and the USA.

Despite these recent breakthroughs, the closed-loop controllers trialed clinically have only utilized a small pool of relatively simple feedback methods, namely bang-bang control (e.g. \cite{little2013}), proportional and proportional-integral control (e.g. \cite{rosa2015,schmidt2024}), and dual-threshold control (e.g. \cite{velisar2019}). The notion of feedback, or closed-loop, which is fundamental to automatic control (see, e.g., \cite{astrom}) and thus to systems biology (see, e.g., \cite{alon,delvecchio}), remains underdeveloped in neuroscience, as evidenced by a number of recent publications (see, e.g., \cite{acharya,closedloop}).
Thus the sad observation \cite{deschutter} on the gap between systems biology and neuroscience, made almost 20 years ago, remains valid today.
The amazing complexity of our brains explains why it is difficult, if not impossible, to obtain a tractable mathematical model of the system of interest (see, e.g., \cite{levenstein,oleary}), and therefore to use existing model-based closed-loop strategies. 

Previously examined closed-loop methods for DBS for PD bypass this difficulty by not utilizing an internal model but relying only on the system outputs to adjust the stimulation signal.
The most common system output for closed-loop control of DBS for PD examined to date is the beta activity (13-30 Hz) in the subthalamic nucleus (STN), which is also a key stimulation target.
The ability to perform recording and stimulation at the same site is attractive, as it allows a design using only a single electrode lead.
The STN beta activity is associated with bradykinetic and rigid symptoms of Parkinson's disease and its suppression (using either medication or DBS) is correlated with improvement in clinical scores \cite{kuhn2006}.

These features make the STN beta activity an attractive biomarker for closed-loop DBS in PD.
However, since the origin of pathological beta and the mechanism of DBS action are still debated, designing a controller that disrupts only pathological beta activity remains a challenge.
Computational efforts in that direction have been made using an adaptive version of proportional control, in which the delivered stimulation intensity is proportional to beta activity through a gain that self-adjusts to various factors such as electrode impedance or disease evolution \cite{fleming}.
While this adaptive proportional stimulation remains silent when no beta activity is detected, it still has an influence on non-targeted brain waves when they superimpose on the pathological ones.

In this context, the technique of model-free control (MFC) \cite{mfc1,mfc2}, which has been recently shown to successfully suppress seizure-like activity in a computational model of epilepsy \cite{join2024epilepsy}, presents a promising avenue to bring more advanced control-theoretical methods into the field of neurostimulation and achieve selective suppression of pathological brain activity, even in the absence of a global model for beta generation and suppression.
In addition to successful engineering applications,\footnote{See numerous references in \cite{mfc1,mfc2}, and a choice of recent ones \cite{coskun,he,kang,la,park,scherer,yaha} among many others. Comparisons \cite{madrid,michel} with other closed-loop control strategies, applied to autonomous vehicles and wind turbines, conclude that MFC outperforms industry standard strategies in a range of metrics.} MFC has already been demonstrated in different applications in biomedicine and bioengineering \cite{bara,anesth,diab,ventil}.
Furthermore, the corresponding controller has been shown to be easily implementable on small and inexpensive programmable devices \cite{chaxel}, which is an important step towards practical implementations of the presented theoretical methods.

This paper is organized as follows.
In Sect. \ref{I}, model-free control with intelligent proportional (iP) controller is introduced.
A brief overview of the differences between the iP and the classic proportional-integral controllers (PI), the most commonly used controllers in engineering, is provided in Sect. \ref{PIiP}.
Sect. \ref{II} introduces the computational model of parkinsonian brain activity, used to test the proposed control methods. 
Sect. \ref{III} presents the results of numerical experiments, illustrating how the iP controller can be used to selectively disrupt pathological beta oscillations in a mean-field type model of Parkinson's disease.
Finally, concluding remarks are presented in Sect. \ref{conclusions}.

\section{Model-free control}\label{I}

\subsection{Ultra-local model}\label{loop}
It has been demonstrated \cite{mfc1} (see \cite{bara} for a more fathomable presentation) that under quite weak assumptions any single-input single-output (SISO) system may be approximated by an \emph{ultra-local model} which needs to be continuously updated:
\begin{align}\label{eq-1}
y^{(\nu)} = F + \alpha u,
\end{align}
where:
\begin{itemize}
\item the control and output variables are respectively $u$ and $y$;
\item the derivation order $\nu$ may be often set to $1$; 
\item the constant $\alpha \in \mathbb{R}$, which is fixed in such a way that $\alpha u$ and
$y^{(\nu)}$ are of the same order of magnitude, does not need to be precisely computed;
\item $F$ subsumes not only the system dynamics but also external disturbances.
\end{itemize}

The ultra-local model \eqref{eq-1} is a linear approximation of the full system dynamics with its external disturbances in a small neighborhood of the operating point.
It can be therefore used to design feedback controllers, taking into account the full knowledge of the local dynamics of the system.
However, the ultra-local nature of this description makes it necessary to continuously update this model by computing a real-time estimate $F_{\rm{est}}$ of $F$ to account for changes in the operating point as well as in the external disturbances. Following the procedure described in detail in \cite{mfc1}, this estimate $F_{\rm{est}}$
can be computed for $\nu = 1$ as\footnote{See \cite{mfc2} for the case $\nu = 2$, which is being used for epilepsy \cite{join2024epilepsy}.}
\begin{equation}\label{estim}
F_{\rm{est}}(t)  =-\frac{6}{\tau^3}\int_{t-\tau}^t \left\lbrack (\tau -2\sigma)y(\sigma)+ \alpha\sigma(\tau -\sigma)u(\sigma) \right\rbrack d\sigma,
\end{equation}
where $\tau > 0$ may be quite small. This quantity can be computed based on the sole knowledge of the applied input $u$ and the measured output $y$.
In practice, the integral in (\ref{estim}) is usually replaced by a digital filter.

\subsection{Intelligent proportional controllers}

For $\nu = 1$, we can define the \emph{intelligent proportional controller} (\emph{iP}) \cite{mfc1} as
\begin{equation}\label{ip}
    u = -\frac{F_{\rm{est}} - \dot{y}^\ast +K e }{\alpha},
\end{equation}
where 
\begin{itemize}
    \item $y^\star$ is the reference trajectory, 
    \item $e = y - y^\star$ is the tracking error, 
    \item $K \in \mathbb{R}$ is the tuning proportional gain. 
\end{itemize}
Substituting \eqref{ip} into \eqref{eq-1}, we get
\begin{align*}
\dot{y} &= F - \alpha \frac{1}{\alpha} \left( F_{\rm{est}} - \dot{y}^\ast +K e \right).
\end{align*}
This gives $
\dot{y} - \dot{y}^\ast = F -F_{\rm{est}} -K e$,
which leads to the following expression of the tracking error
$$
\dot{e} =- K e + F - F_{\rm{est}}.
$$
For $K> 0$, we thus ensure that if the estimate $F_{\rm{est}}$ of $F$ is good (i.e. $F - F_{\rm{est}} \approx 0$), then $\lim_{t \to \infty} e(t) \approx 0$: the output $y$ approximately reaches the reference $y^*$. 

\begin{remark}
Since the signals used to estimate the ultra-local model are corrupted by unavoidable measurement noise (see, e.g., \cite{tagawa}), it is crucial that the estimate $F_{\rm{est}}$ can be performed under noisy measurement conditions.
This issue is generally addressed by statistical and/or probabilistic tools that are difficult to model in most real-life situations.
Following \cite{noise}, the noise is related via a deep result, due to Cartier and Perrin \cite{cartier}, to quick fluctuations around $0$.
Such a fluctuation, which is expressed in the language of \emph{nonstandard analysis}, is
a Lebesgue-integrable real-valued time function ${\mathfrak{f}}(t)$ whose integral is infinitesimal over any   bounded time interval. 
Therefore, noise is attenuated thanks to the integral in Formula \eqref{estim}.
This point of view, which bypasses all probabilistic and statistical considerations, has been confirmed not only in model-free control (see previous references) but also in signal processing (see, e.g., \cite{beltran,mboup,othmane}).
\end{remark}

\subsection{Comparison between sampled PI and iP controllers}\label{PIiP}
The proportional-integral (PI) controller is without any doubt the most important controller in control engineering \cite{astrom}.
However, the issue of tuning the PI controllers remains an unsolved challenge when the model dynamics are not known.
Here we show that the iP controller, which can account for those unmodelled dynamics, is equivalent to a PI controller in a discrete-time case.

Consider the continuous-time {\em proportional-integral controller} (\emph{PI}):
\begin{equation}\label{cpi}
  u (t) = k_p e(t) + k_i \int_0^t e(\tau) d\tau,
\end{equation}
where $k_p$ and $k_i$ denote the proportional and integral gains respectively.

Differentiating both sides of \eqref{cpi} with respect to time, we obtain the ``velocity form'' of the PI controller:
\begin{equation*}\label{cpid}
\dot{u} (t) = k_p \dot{e}(t) + k_i e(t).
\end{equation*}
Sampling this controller with a sampling interval $h$ yields the following discrete-time form
\begin{equation}\label{pidisc}
\dfrac{u(t) - u(t - h)}{h} =  k_p  \left(\dfrac{e(t) - e(t -
h)}{h}\right) + k_i  e(t).
\end{equation}

By replacing $\dot{y}$ and $u$ in \eqref{eq-1} and \eqref{ip} by their sampled versions, we obtain the following expression for $F$:
$$F = \frac{y(t) - y(t-h)}{h} - \alpha u (t-h)$$
Using this expression for $F_{\rm{est}}$ in \eqref{ip}, we get
\begin{equation}
\label{eqDiscr_i-POne} u (t) = u (t - h) - \frac{e(t) -
e(t-h)}{h\alpha} + \dfrac{K}{\alpha}\, e(t)
\end{equation}
Thus, the control laws expressions \eqref{pidisc} and
\eqref{eqDiscr_i-POne} become identical for the following choices of control gains:
\begin{align}
\label{eqPI_i-P_corresp} k_p &= - \dfrac{1}{\alpha h}, \quad k_i =
\dfrac{K}{\alpha h}.
\end{align}
It should be emphasized that this relationship between iP and PI controllers does not hold in
continuous-time. This equivalence is strictly related to
time sampling, {\em i.e.}, to computer implementation, as
demonstrated by taking $h \downarrow 0$ in 
\eqref{eqPI_i-P_corresp}.

\section{Model description}
\label{II}
The main strength of the control strategy presented in the above section is that is does not rely on a detailed model of the dynamics involved. In view of the complexity of the brain dynamics involved and inherent uncertainties in biological models, this advantage is particularly well suited to counteract Parkinsonian brain oscillations.

To assess the performance of intelligent proportional control in the context of DBS, we consider the following model which has already been extensively used to model pathological oscillations and their suppression in PD  \cite{fleming,nevadoholgado,pasillaslepine,pavlides}.
This model, originally introduced in \cite{nevadoholgado}, describes the activity of two nuclei of the basal ganglia: the subthalamic nucleus (STN) and the external globus pallidus (GPe). It reads
\begin{subequations}\label{Model}
\begin{align}
\tau_1\dot x_1(t)&=-x_1(t)+S_1\Big( c_{11}x_1(t-\delta_{11})-c_{12}x_2(t-\delta_{12})+b_1(p(t)+u_1(t)) \Big)\\
\tau_2\dot x_2(t)&=-x_2(t)+S_2\Big( c_{21}x_1(t-\delta_{21})-c_{22}x_2(t-\delta_{22})+b_2u_2(t) \Big),
\end{align}
\end{subequations}
where $x_1(t)$ and $x_2(t)$ represent the instantaneous firing rate of STN and GPe respectively.
The gains $c_{ij} \geq 0$ represent the strength of the synaptic connection from population $j$ to population $i$.
The two structures are reciprocally interconnected in a bidirectional way: STN exerts an excitatory influence on GPe whereas GPe provides inhibition of the STN (as indicated by the negative sign in front of $c_{12}$ and $c_{22}$).
The non-instantaneous propagation of action potentials along the axons is captured by the delays $\delta_{ij} \geq 0$. $\tau_1, \tau_2 \geq 0$ are time constants.
The functions $S_1, S_2: \mathbb{R} \to \mathbb{R}$ are the activation functions, encoding the nonlinear input-output relationship of STN and GPe, respectively.
The system has three inputs: $u_1$ representing the control input (stimulation electrode present only in the STN), a perturbation $p$ representing non-modelled brain signals to STN (primarily from cortex), and $u_2$ representing the exogenous inputs to GPe (primarily from striatum).
These inputs are applied to the subsystems with gains $b_1, b_2 \geq 0$.

With appropriate choice of parameters (studied in detail in \cite{nevadoholgado,pasillaslepine,pavlides}), this system exhibits sustained oscillations in the beta range as a result of instability caused by strong synaptic coupling between the two populations. The parameter values used in this paper are $c_{11}=0$, $c_{12}=3$, $c_{21}=10$, $c_{22}=0.9$, $b_1=5$, $b_2=139.4$, $\tau_1=6$\,ms, $\tau_2=14$\,ms, $\delta_{11}=\delta_{22}=4$\,ms and $\delta_{12}=\delta_{21}=6$\,ms.
The activation functions $S_i$ are defined as
\begin{equation*}
    S_i(x) = \frac{m_i b_i}{b_i + \exp(-4x/m_i)(m_i - b_i)},
\end{equation*}
where $m_1=300$, $b_1=17$, $m_2=400$, $b_2=75$.

The model \eqref{Model} constitutes a rough abstraction of the neuronal dynamics involved.
It obviously omits modelling the inherent heterogeneity and stochasticity of the individual neurons behavior.
Yet, it is well suited to capture dynamics taking place at the population scale, such as brain oscillations.
It is also worth stressing that this model serves purely for validation purposes: the model-free control strategy employed here, as its name indicates, assumes no knowledge at all of the underlying dynamics.

\section{Numerical experiments}
\label{III}

We now proceed to numerical investigations on the population model \eqref{Model}, to validate the efficiency of the intelligent proportional control at disrupting pathological oscillations, assess its robustness to additive noise, and investigate its ability to leave non-pathological activity unaltered.

\begin{figure}[!ht]
\centering
{\epsfig{figure=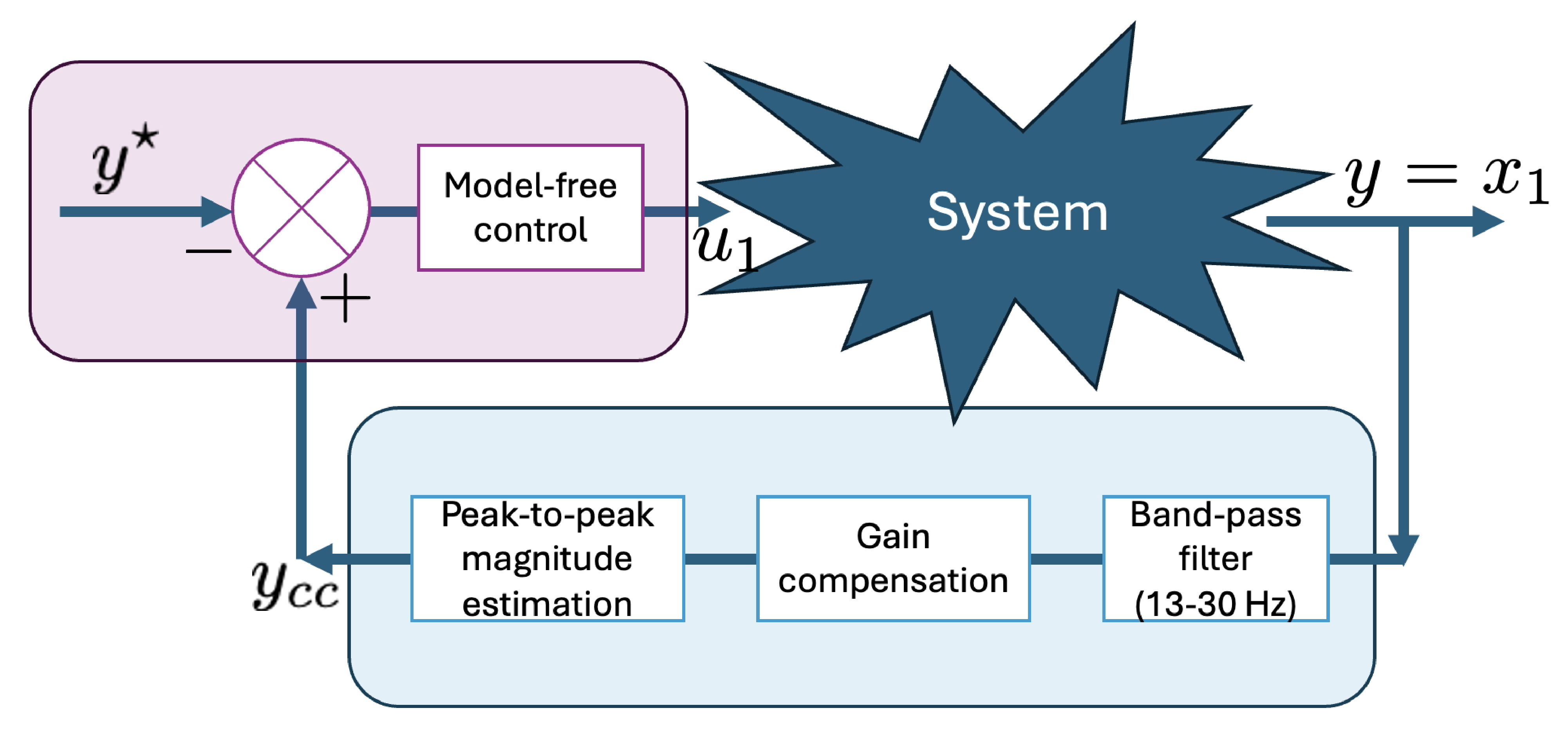,width=1.05\textwidth}}\caption{Control scheme.}\label{CS}
\end{figure}

The control scheme used to disrupt pathological oscillations in \eqref{Model} is depicted in Figure \ref{CS}.
The measured output $y = x_1$ represents the instantaneous firing rate of the STN.
The biomarker of interest, $y_{cc}$, is extracted in a three-step procedure:
\begin{enumerate}
\item FIR band-pass filter is used to extract the pathological activity in the beta \hbox{(13 -- 30\,Hz)} range,
\item gain compensation is applied to correct the non-unitary gain of the band-pass filter,
\item peak-to-peak amplitude of the resulting signal is obtained by taking the difference between the maximum and minimum value of the signal over a $1/13$\,s sliding window (corresponding to the maximum period of a beta wave). 
\end{enumerate} 

Denoting by $y_{cc}$ the obtained gain-compensated peak-to-peak amplitude of the beta wave oscillations, we rely on the following ultra-local model (see Section \ref{I}):
\begin{equation}
    \dot y_{cc}=F+\alpha u_1.
\end{equation}
Accordingly, the closed-loop control reads
\begin{align}\label{eq-2}
u_1=\frac{1}{\alpha}\left(\dot y^\star -F_{\rm{est}} + K e\right),
\end{align}
where $e=y^\star-y_{cc}$ and where we have picked $\alpha=50$ and $K=5$.

As we seek to reduce $y_{cc}$ to $0$, we might be tempted to use the reference as $y^\star=0$.
However, our experience has shown that it is preferable to use a very low setpoint $y^\star=\varepsilon > 0$. 
This is done to guarantee low-amplitude control at all times, as setting $y^\star$ to 0 may lead to the controller output getting stuck at a value that is successfully suppressing the oscillations but is unnecessarily high.

In all cases, $y^\star$ is constant and $\dot y^\star=0$.
The control law \eqref{eq-2} thus boils down to 
$$u_1=\frac{1}{\alpha}\left( -F_{\rm{est}} + K e\right).$$

\begin{figure*}[!ht]
\centering
\subfigure[\footnotesize States]
{\epsfig{figure=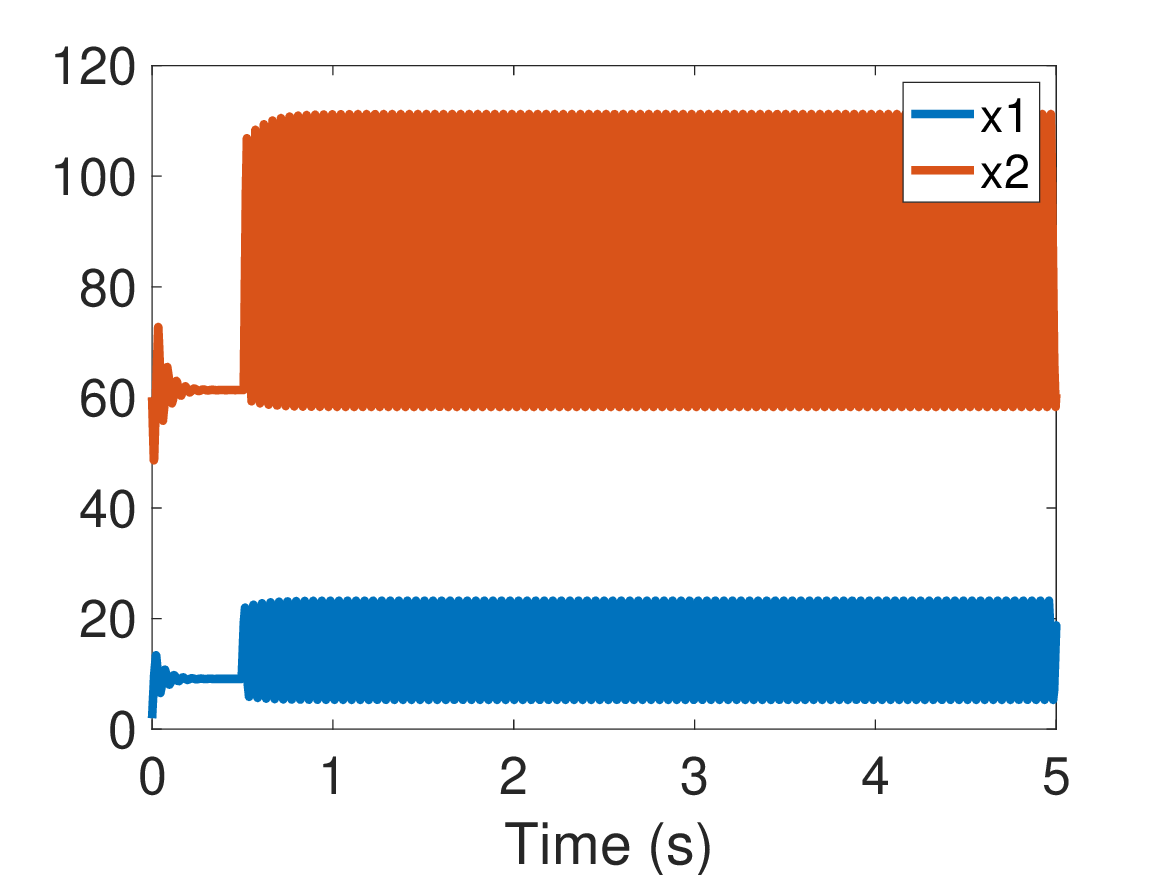,width=0.35\textwidth}}
\subfigure[\footnotesize Filtered output]
{\epsfig{figure=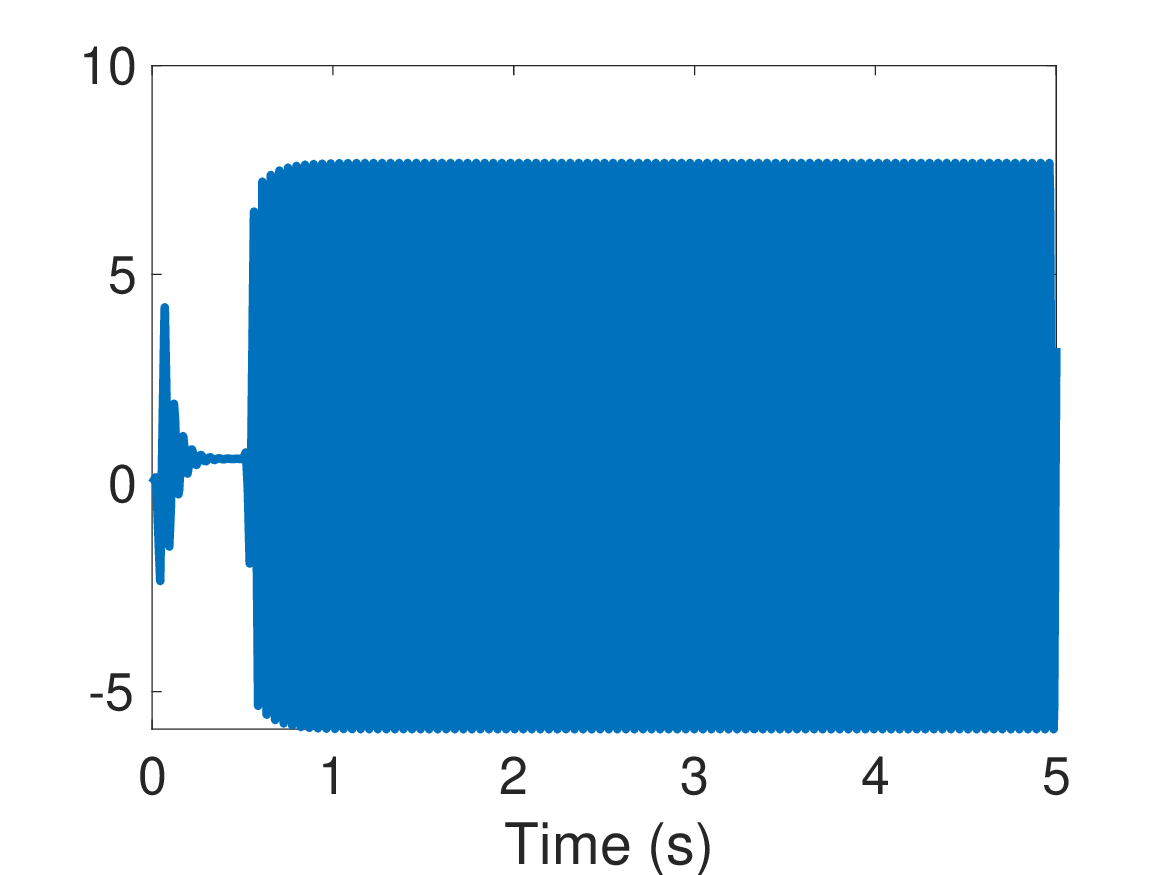,width=0.35\textwidth}}
\subfigure[\footnotesize Magnitude estimation]
{\epsfig{figure=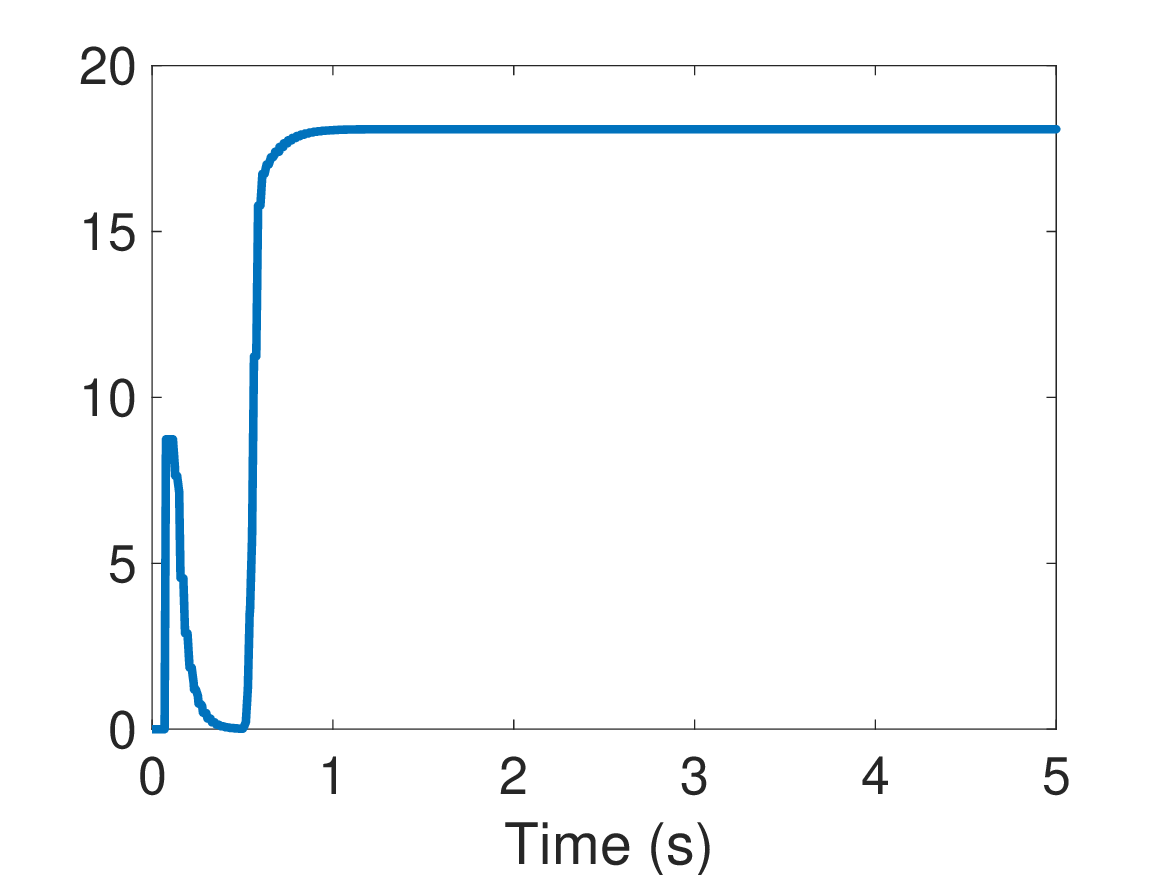,width=0.35\textwidth}}
\newline
\subfigure[\footnotesize States]
{\epsfig{figure=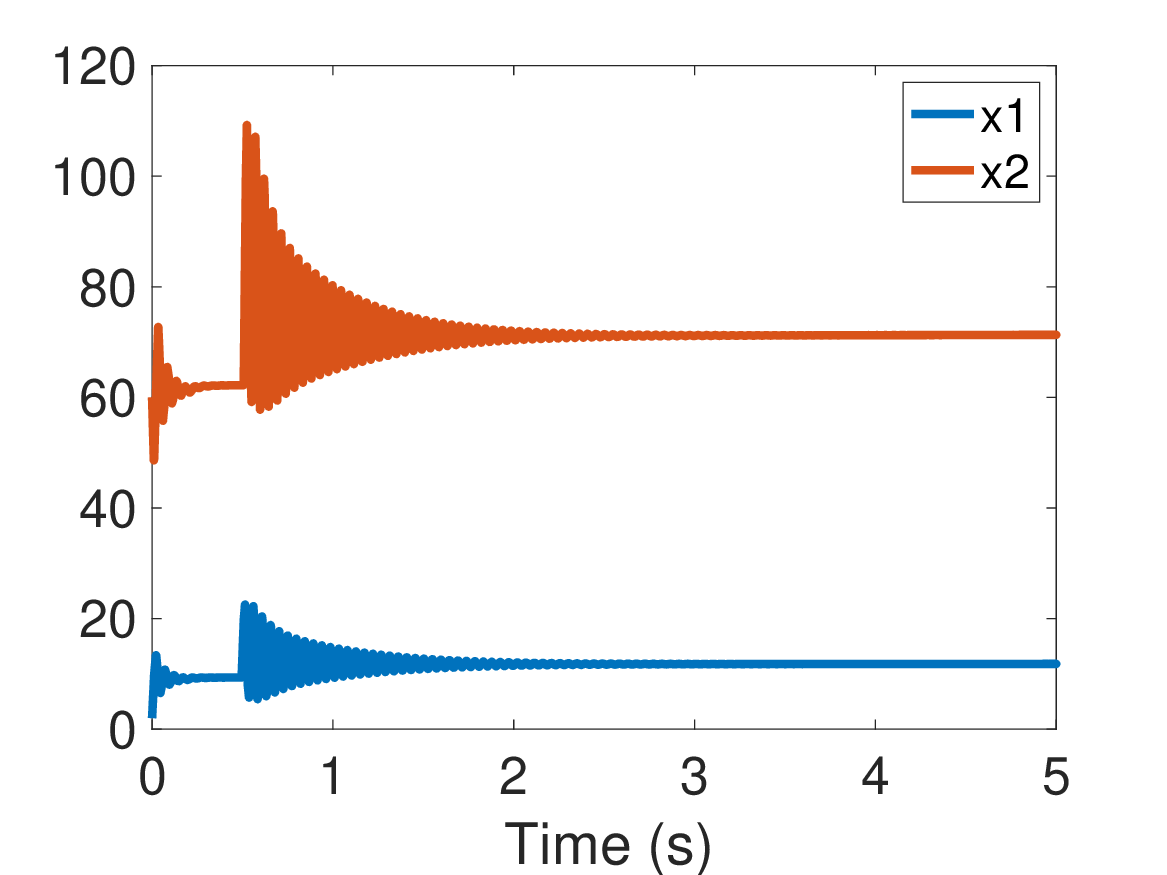,width=0.35\textwidth}}
\subfigure[\footnotesize Control input ]
{\epsfig{figure=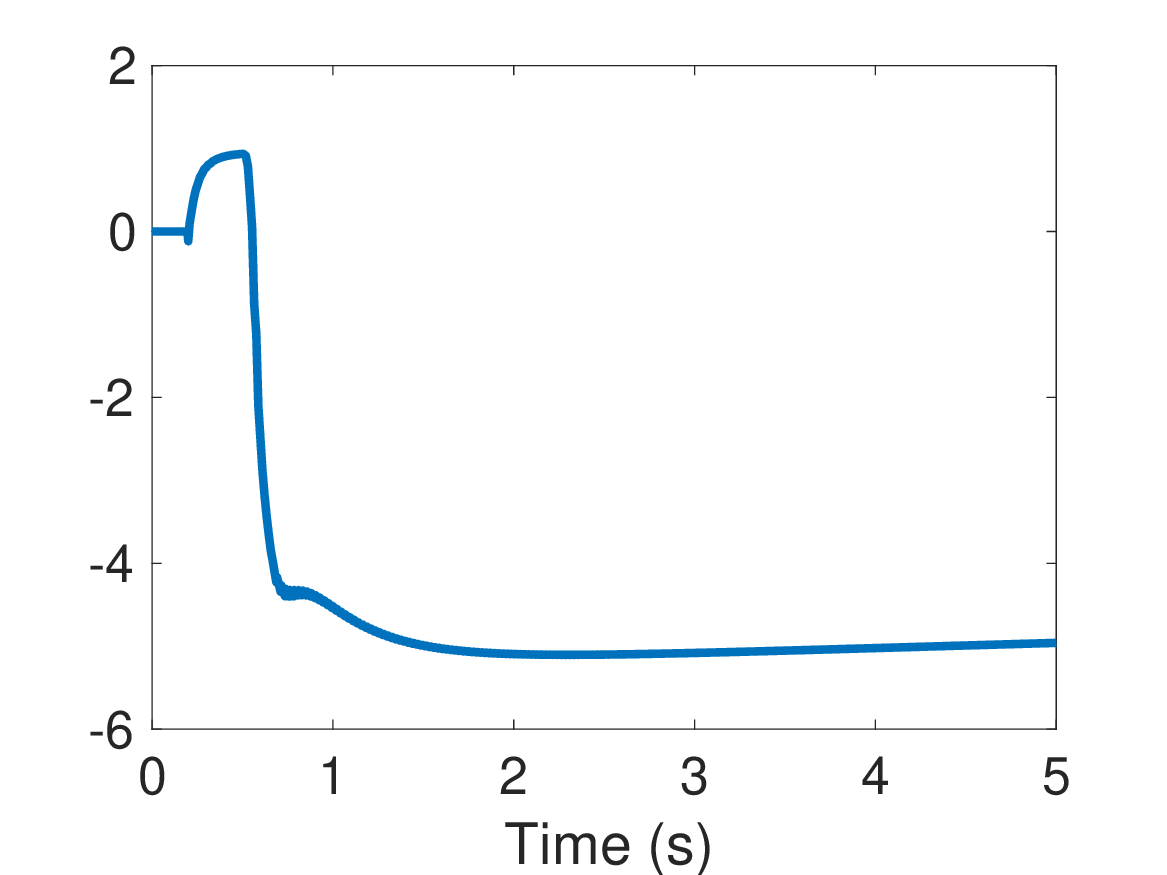,width=0.35\textwidth}}
\subfigure[\footnotesize Setpoint (- -) and magnitude estimation (--)]
{\epsfig{figure=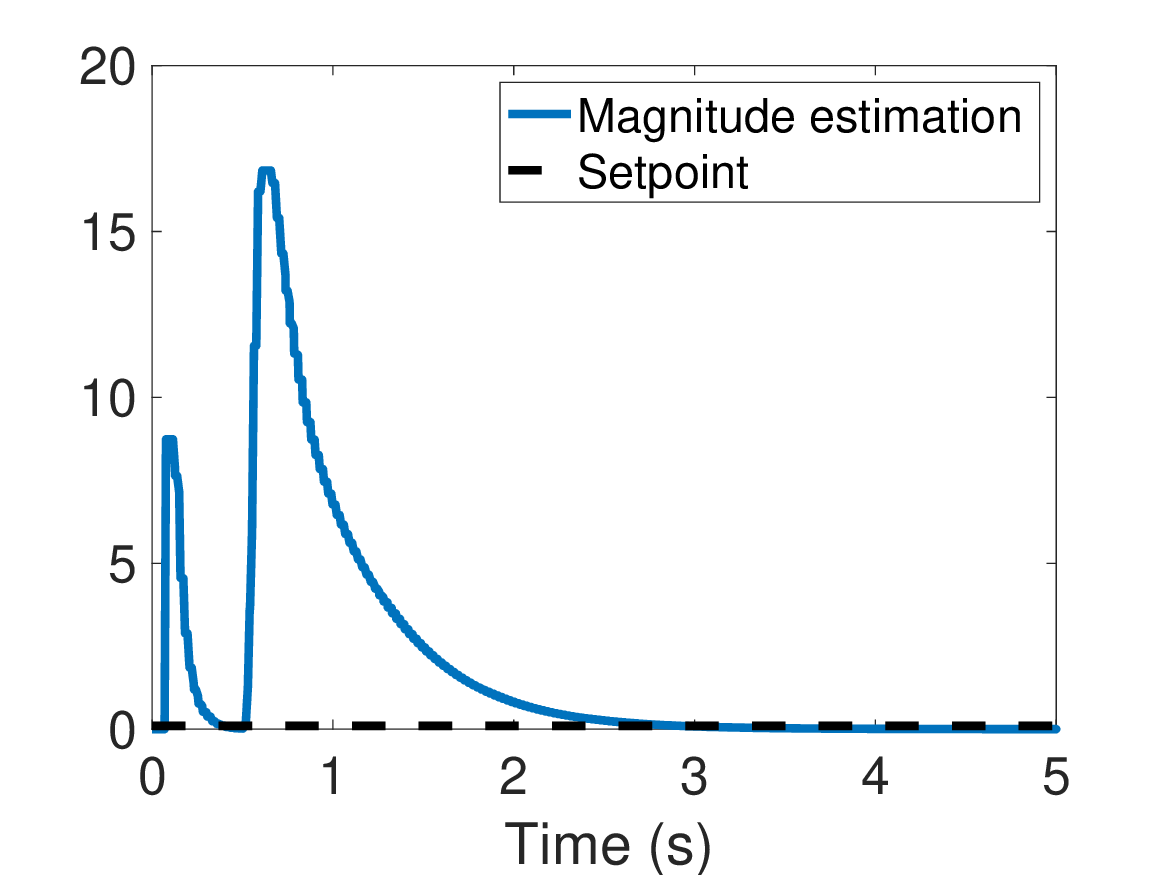,width=0.35\textwidth}}
\caption{Results for Scenario 1. \textbf{Top row}: Open loop. \textbf{Bottom row}: Closed-loop with stimulation.}\label{S1O}
\end{figure*}
\begin{figure*}[!ht]
\centering
\subfigure[\footnotesize States]
{\epsfig{figure=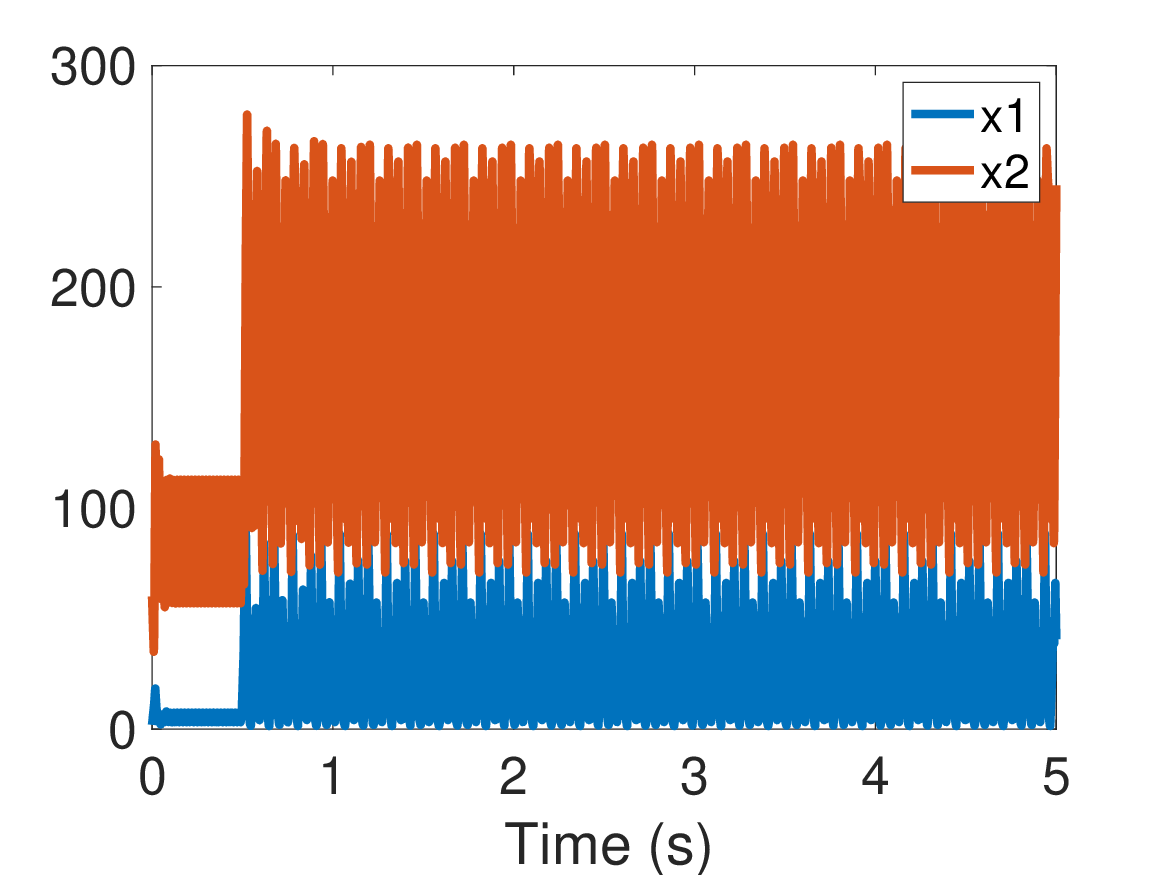,width=0.35\textwidth}}
\subfigure[\footnotesize Filtered output ]
{\epsfig{figure=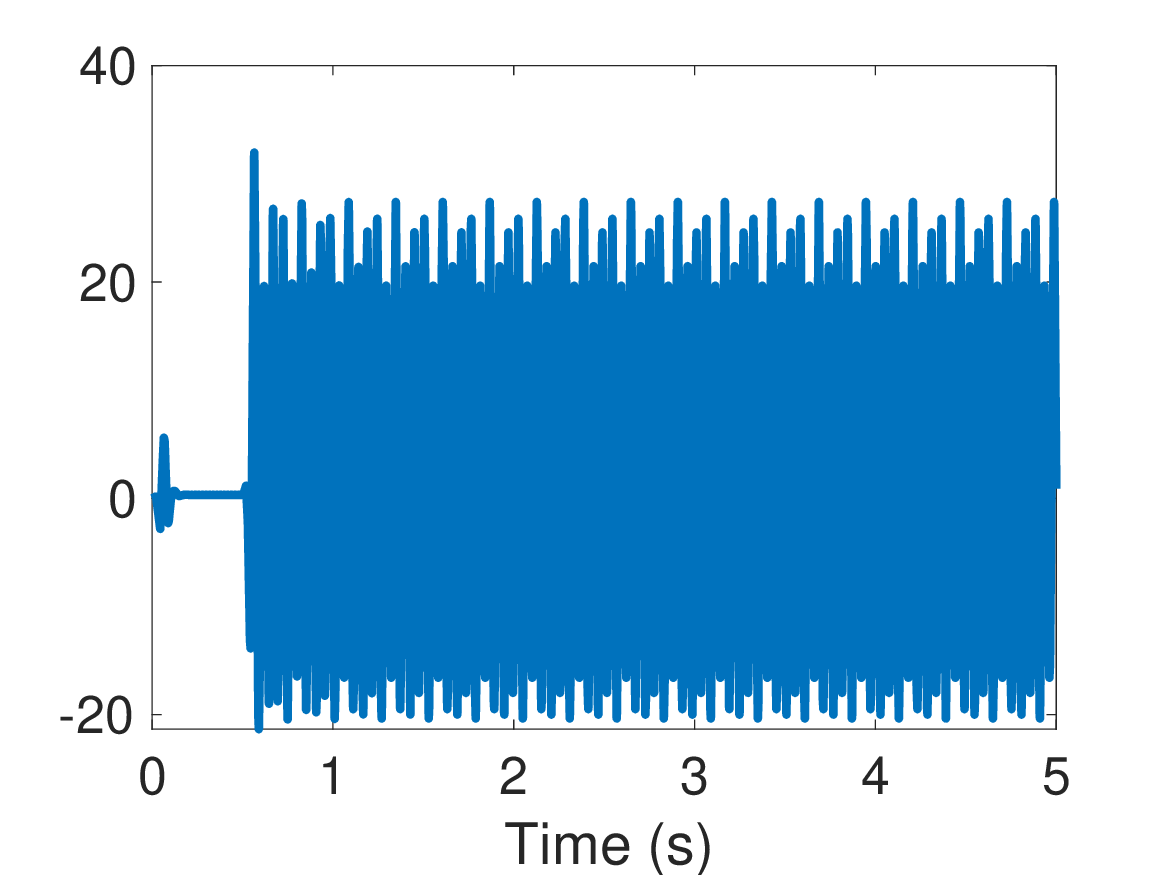,width=0.35\textwidth}}
\subfigure[\footnotesize Magnitude estimation]
{\epsfig{figure=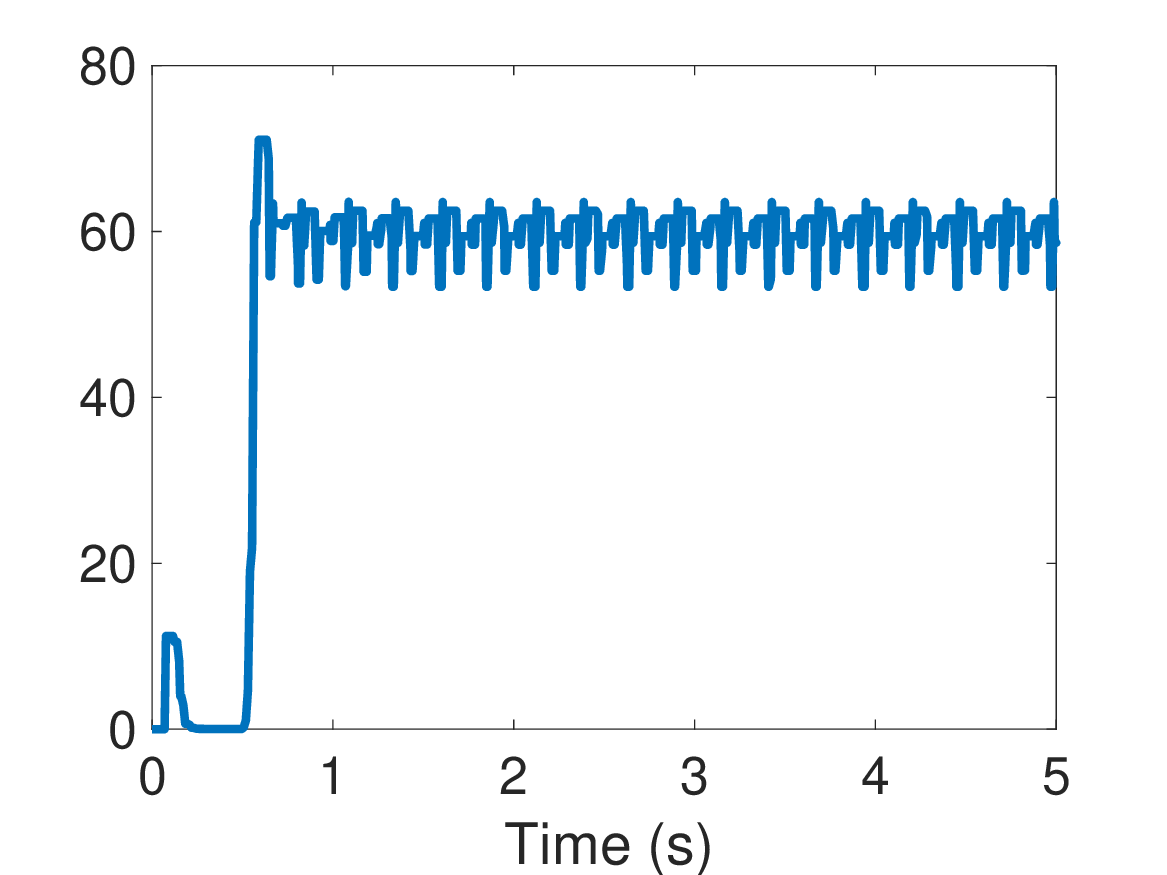,width=0.35\textwidth}}
\newline
\subfigure[\footnotesize States]
{\epsfig{figure=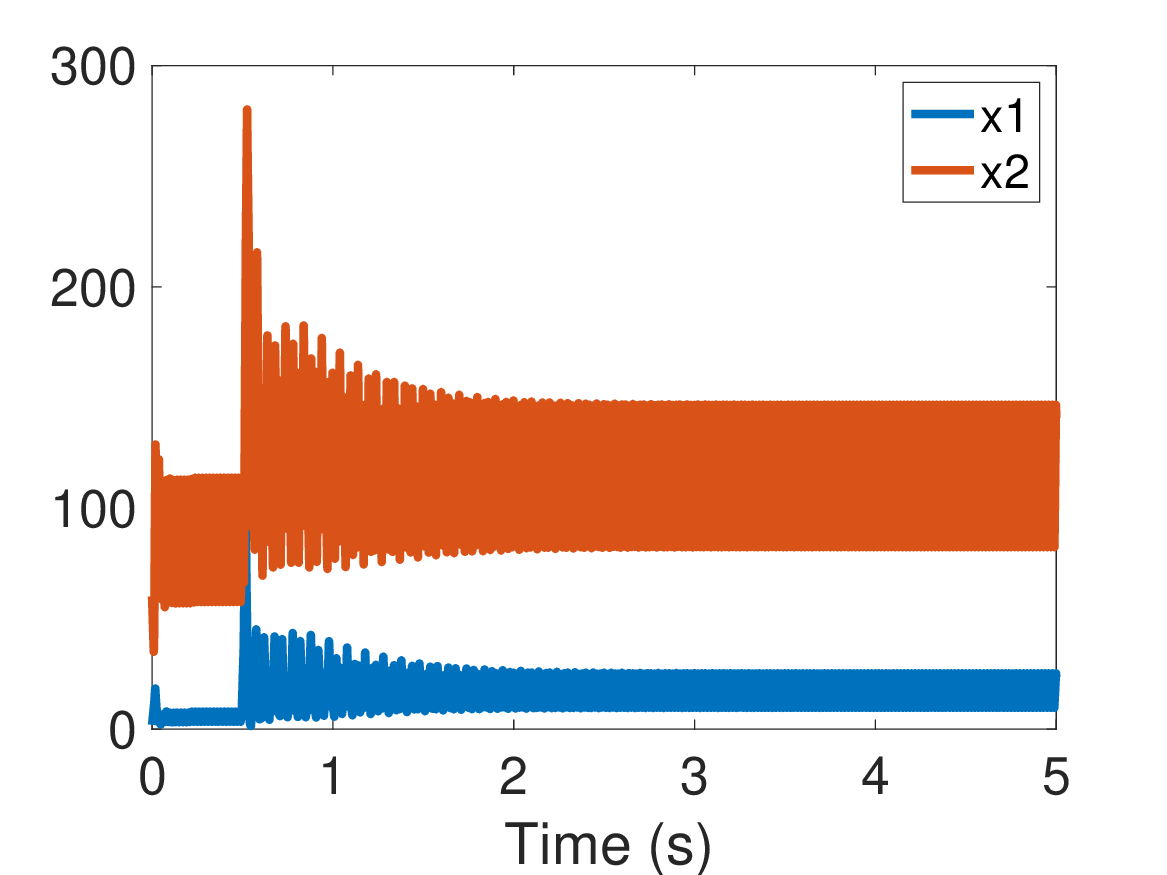,width=0.35\textwidth}}
\subfigure[\footnotesize Control input ]
{\epsfig{figure=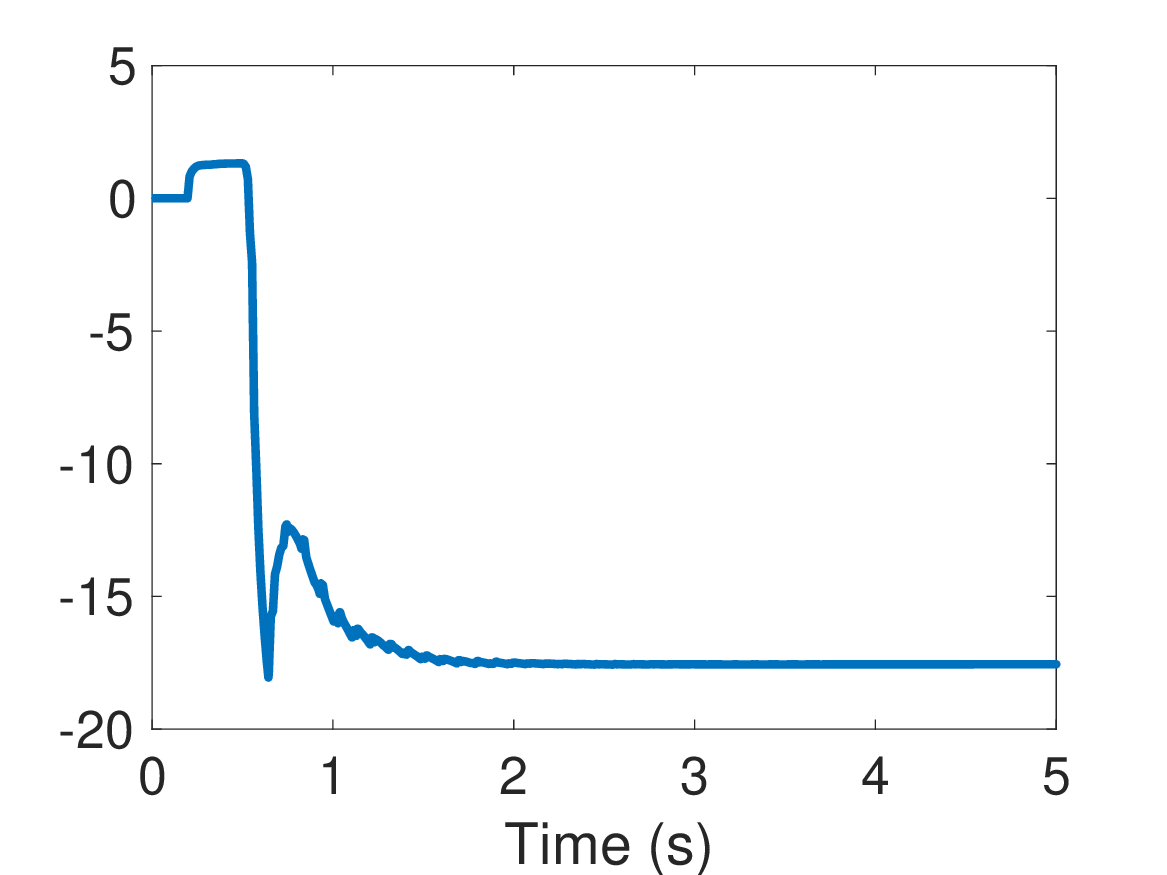,width=0.35\textwidth}}
\subfigure[\footnotesize Setpoint (- -) and magnitude estimation (--)]
{\epsfig{figure=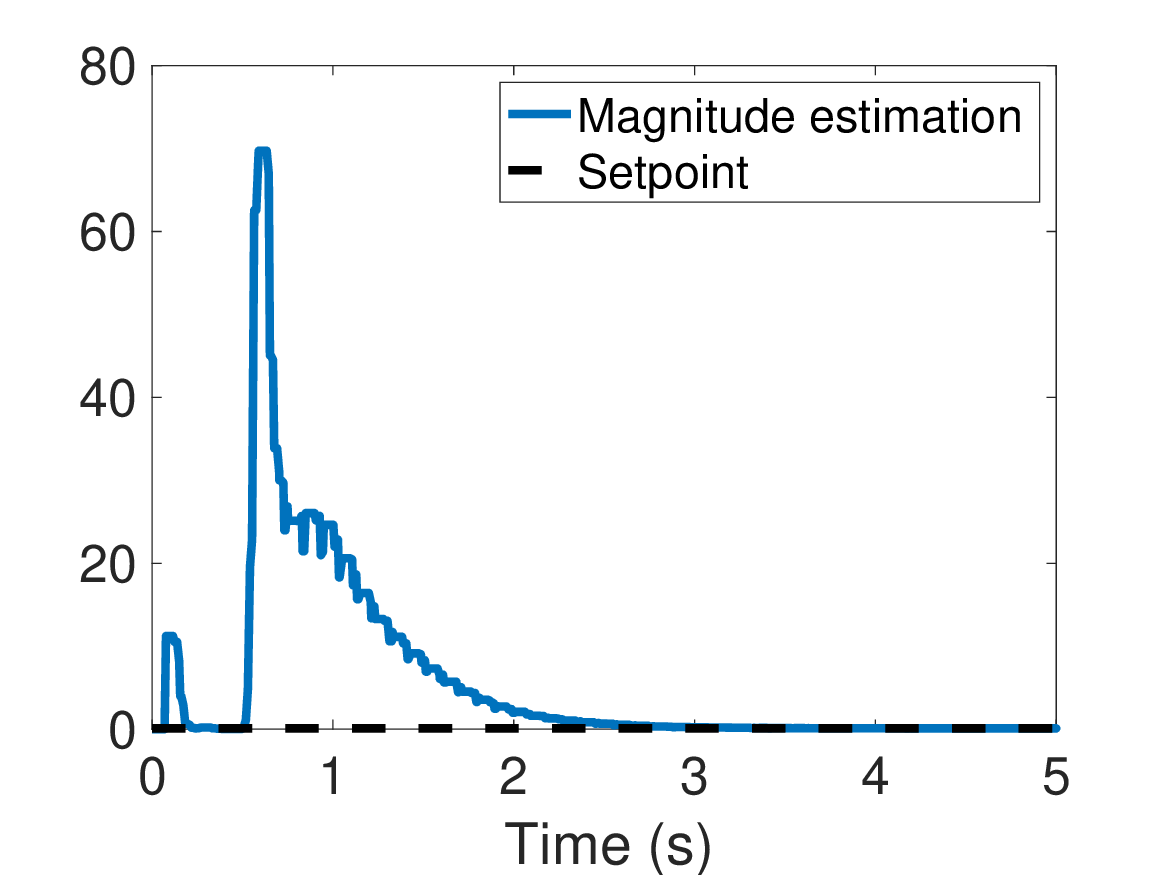,width=0.35\textwidth}}
\caption{Results for Scenario 2. \textbf{Top row}: Open loop. \textbf{Bottom row}: Closed-loop with stimulation.}\label{S2O}
\end{figure*}
\begin{figure*}[!ht]
\centering
\subfigure[\footnotesize States]
{\epsfig{figure=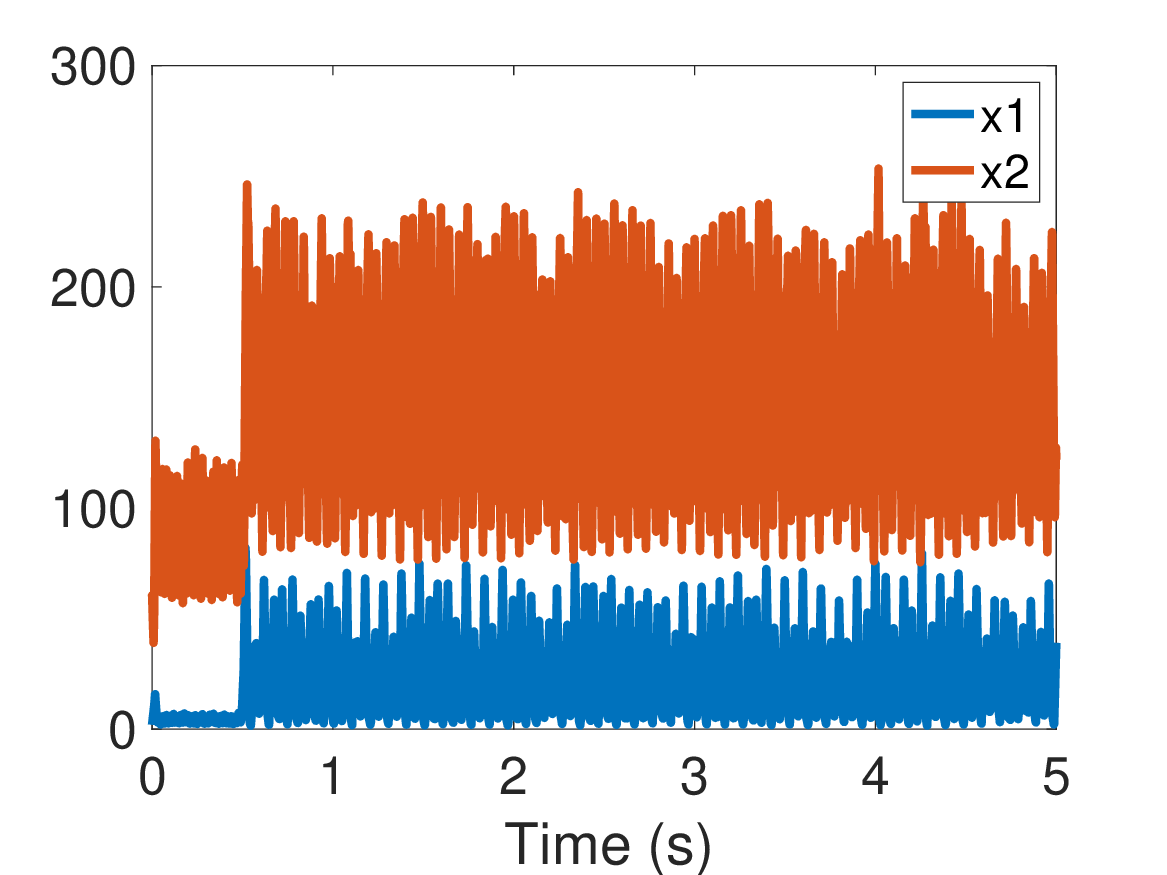,width=0.35\textwidth}}
\subfigure[\footnotesize Filtered output ]
{\epsfig{figure=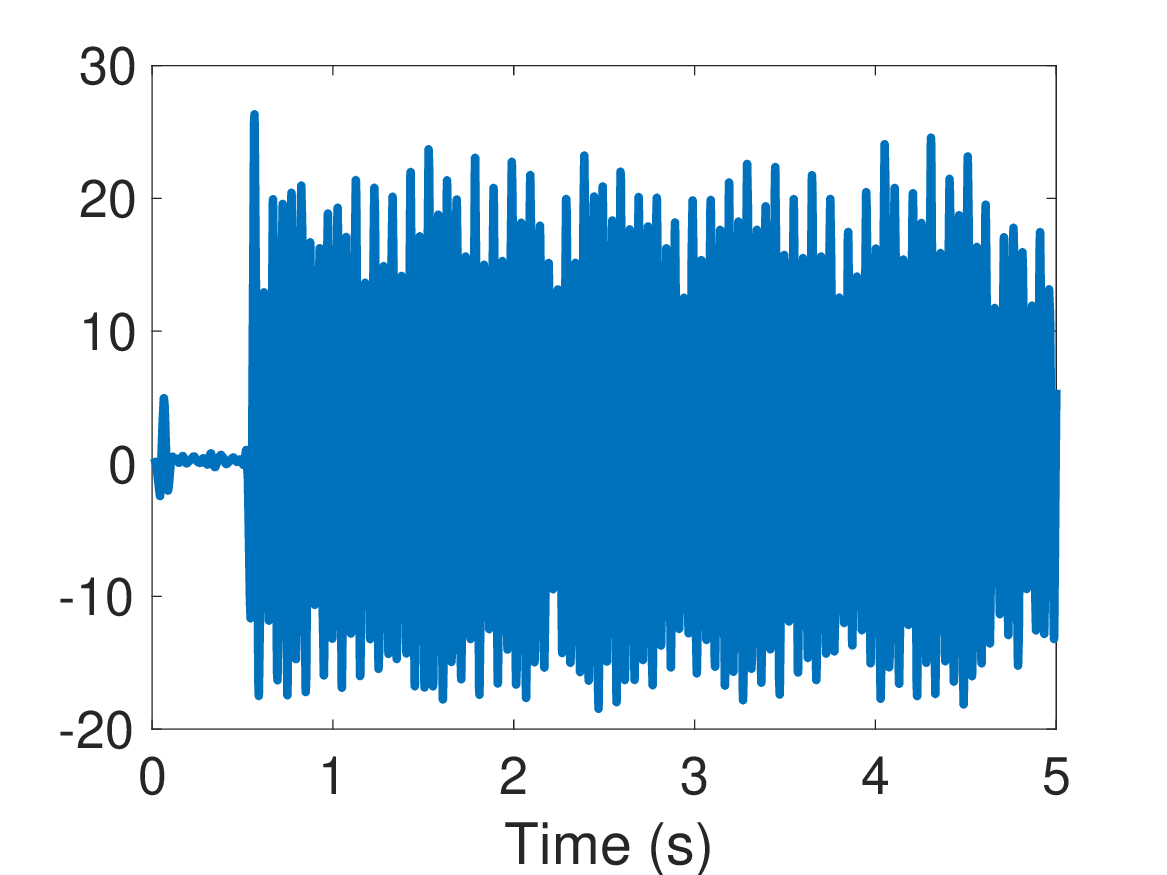,width=0.35\textwidth}}
\subfigure[\footnotesize Magnitude estimation]
{\epsfig{figure=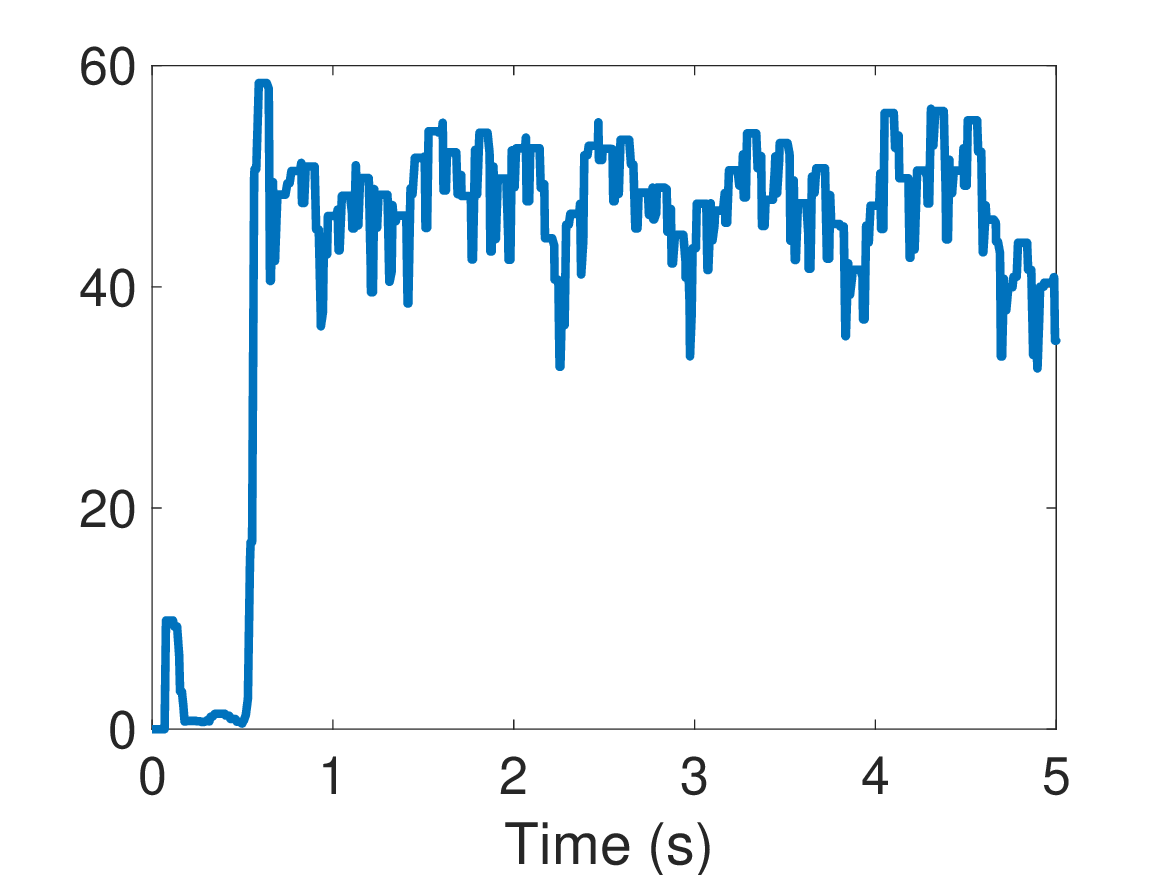,width=0.35\textwidth}}
\newline
\subfigure[\footnotesize States]
{\epsfig{figure=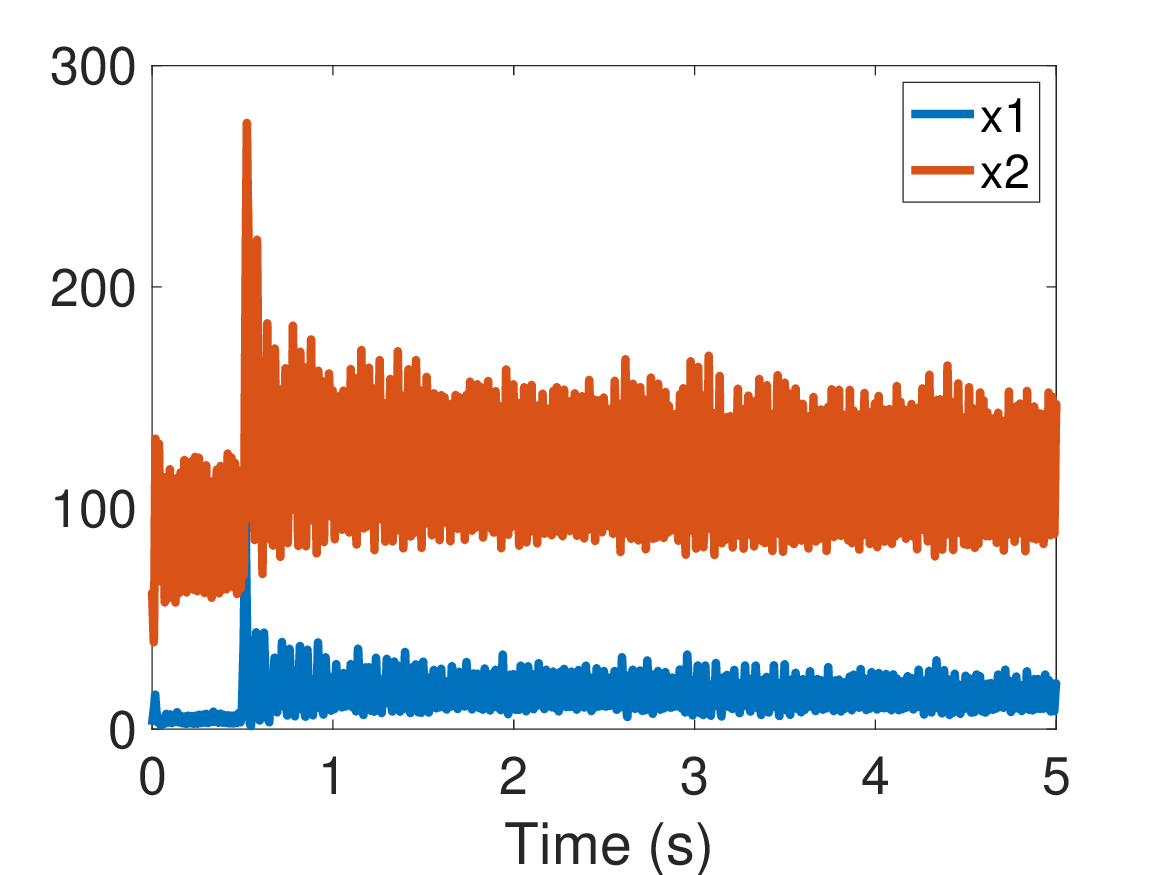,width=0.35\textwidth}}
\subfigure[\footnotesize Control input ]
{\epsfig{figure=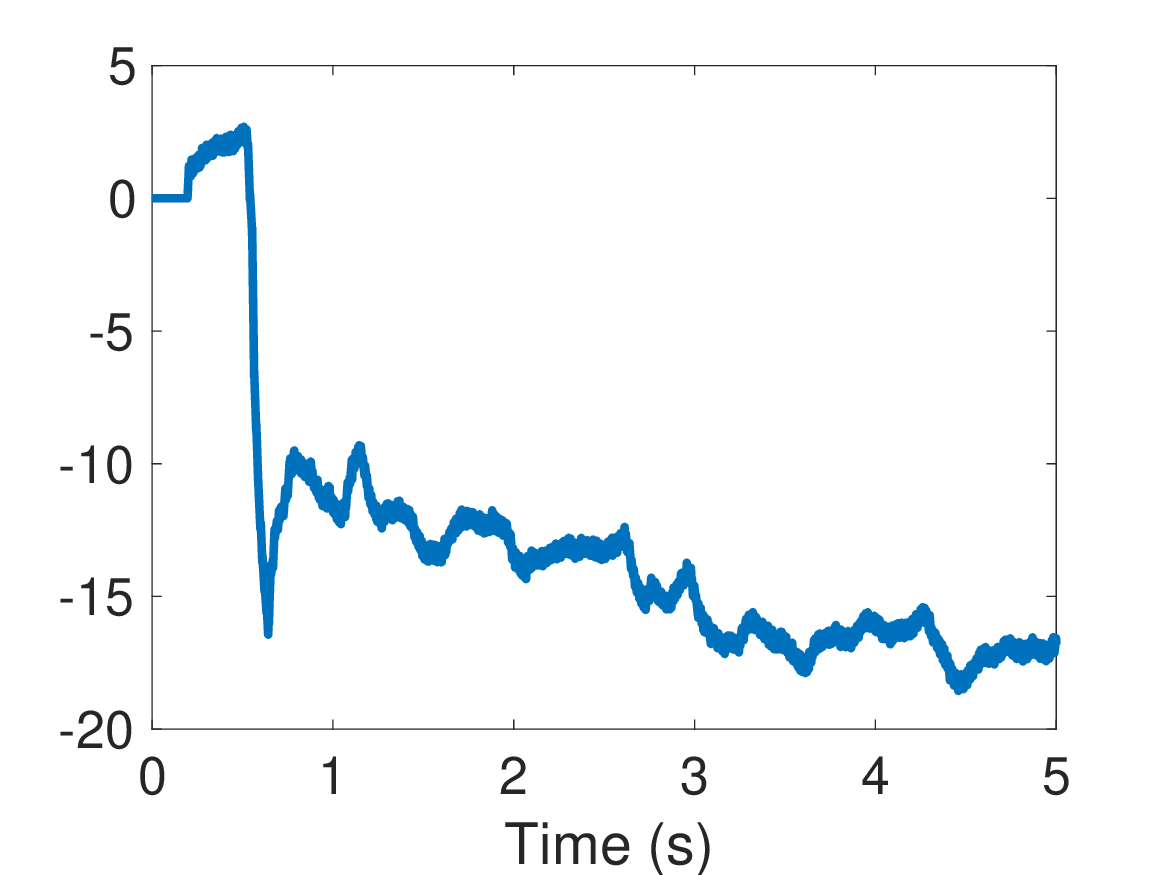,width=0.35\textwidth}}
\subfigure[\footnotesize Setpoint (- -) and magnitude estimation (--)]
{\epsfig{figure=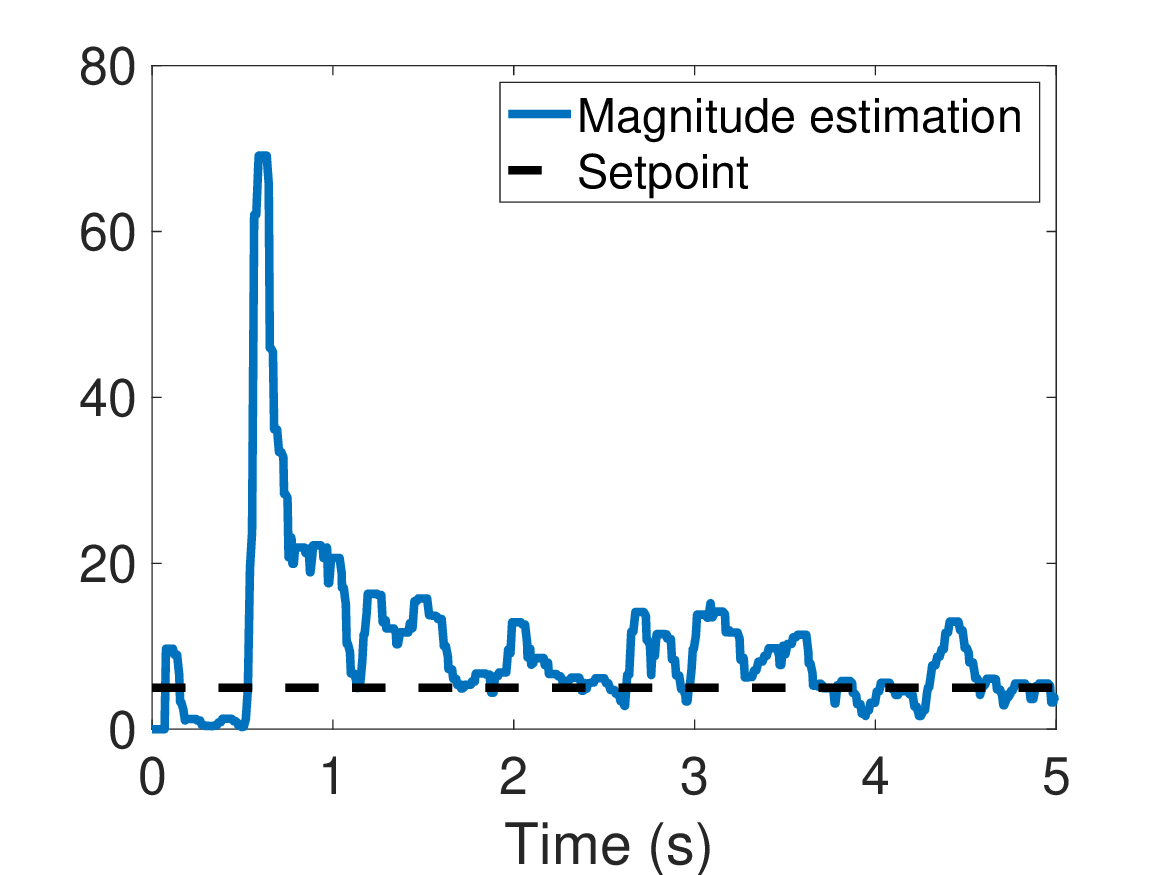,width=0.35\textwidth}}
\caption{Results for Scenario 3. \textbf{Top row}: Open loop. \textbf{Bottom row}: Closed-loop with stimulation.}\label{S3O}
\end{figure*}

Three scenarios illustrate the obtained results:
\begin{itemize}
\item Scenario 1: the cortical input to $x_1$ is $p(t)=27$ for $t\in[0,500]$\,ms then $p(t)=42$ for $t\geq 500$\,ms, whereas the striatal input to $x_2$ is constant: $u_2(t)=60/139.4$ for all $t\geq 0$,
\item Scenario 2: $p(t)=27$ for $t\in[0,500]$\,ms then $p(t)=80$ for $t\geq 500$\,ms and $u_2(t)=60/139.4+2\sin(2\pi50t)$ for all $t\geq 0$,
\item Scenario 3: two normally distributed random samples from $\mathcal{N}(0,1)$ distribution are added to $p(t)$ and $u_2(t)$ from Scenario 2 at each timestep.
\end{itemize}

These three scenarios have specific motivations. In Scenario 1, the cortical input is first fixed to a low value ($p(t)=27$), which does not give rise to sustainded beta oscillations in the network.
The constant value $p(t)=42$ places the operating point in an unstable configuration, which translates into the generation of a stable limit cycle corresponding to sustained beta activity.
In Scenario 2, exogenous oscillatory activity, originating in the striatum, is introduced.
Its frequency (50\,Hz) is in the gamma band, thus outside of the targeted bandwidth.
Finally, Scenario 3 illustrates the controller performance in the presence of noise resulting from surrounding neuronal activity.

These scenarios were simulated using custom Matlab scripts.
The simulation timestep was set at 0.1 ms and Equations \eqref{Model} were integrated using the \texttt{ode45} solver.
At each simulation step, the ultra-local model $F_{est}$ was obtained using a first-order filter implementation of the integral shown in Equation \eqref{estim} with the time constant $\tau$ = 800 ms.

For each of these scenarios, open-loop ($u_1=0$) and closed-loop results are shown in Figures \ref{S1O}, \ref{S2O} and \ref{S3O}, with the open-loop results presented in the top row and closed-loop results in the bottom row.
In all cases, the open-loop filtered output is shown in Figure $\{2,3,4\}$-(b) and the gain-compensated oscillation amplitude estimate $y_{cc}$ is shown in Figure $\{2,3,4\}$-(c) for the open-loop simulation and Figure $\{2,3,4\}$-(f) for closed-loop simulation.
The controller aims to bring the error $e=y^\star-y_{cc}$ close to $0$, which it succeeds in, as measured by the distance separating the two curves shown in Figures $\{2,3,4\}$-(f) (with $y^\star$ indicated by a dashed line).

In the noise-free case (Scenarios 1 and 2), the reference is set to $y^*=0.1$, whereas in Scenario 3 the reference is set to $y^*=5$.
In the latter case, the low setpoint of Scenarios 1 and 2 is unattainable, due to the presence of the noise, which would inevitably lead to saturation of the control at its maximum admissible value, with no benefit for the treatment.

The control input calculated by our algorithm is applied from $t>0.2$\,s to allow the filter and $F$ estimator to operate.
As a consequence, the applied control law has virtually no impact on the transient oscillations observed for $p(t)=27$ (time interval $[0,500]$\,ms on Figure \ref{S1O}-(a) and \ref{S1O}-(d)) as these oscillations are already significantly damped before the simulation is activated.

Comparison of the Figures \ref{S1O}-(c), \ref{S1O}-(f), \ref{S2O}-(c), \ref{S2O}-(f), \ref{S3O}-(c) and \ref{S3O}-(f) shows that the proposed control scheme is able to drastically reduce the amplitude of beta oscillations in all 3 scenarios. 

It is also important to note that during this phase of beta-wave control, the remaining  brain activity is not reduced to zero, as shown in Figures \ref{S1O}-\ref{S2O}-\ref{S3O}-(d): the gamma-band oscillations induced by the additive sine wave in Scenarios 2 and 3 are well preserved.
This suggests that the adopted strategy is a viable candidate for selective disruption of pathological brain oscillations, with limited interference on the non-targeted frequency bands.

Note also that the noise added in Scenario 3 has a notable influence on the closed-loop performance. However, online estimation of $F$ is not severely affected by this additional noise.

\section{Conclusion}
\label{conclusions}

This paper presents preliminary results in the development of model-free control strategies for closed-loop DBS in Parkinson's disease.
We have shown through simulations that, in a nonlinear population model with delays, representing the neural network central to pathological oscillations targeted by closed-loop DBS for PD, it successfully disrupts beta-band activity with limited influence on other exogenous brain waves.
This selective disruption is very appealing as it would result in a more selective treatment in which pro-kinetic activity would be less altered than in open-loop or on-demand DBS.
The online estimation of the local model is another attractive feature, as neuronal population models are inherently uncertain due to their biological nature and the limited measurements available.
Interestingly, our simulations suggest that this approach copes well with the inherent noise involved.

A next step towards clinical translation would be to evaluate this control strategy on a more detailed model of the neural dynamics involved, such as the one developed in \cite{fleming}, which involves additional neuronal populations within the cortico-basal ganglia network and simulates individual  neural activity with conductance-based models, in addition to capturing antidromic effects of DBS.
This model also allows access to clinically relevant measurements such as local field potentials, providing a bridge between network activity and recorded brain signals.


\clearpage 
\begin{thebibliography}{99}

\bibitem{acharya}
Acharya, G, Ruf, S.F., Nozari, E.: Brain modeling for control: A review.
Front. Control. Eng. \textbf{3}, 1046764 (2022)

\bibitem{alon}
Alon, U.: An Introduction to Systems Biology. 2nd edn. CRC Press, Boca Raton, FL (2020).

\bibitem{madrid}
Artu$\tilde{\rm n}$edo, A., Moreno-Gonzalez, M., Villagra, J.: Lateral control for
autonomous vehicles: A comparative evaluation. Annual Rev. Contr. {\bf 57}, 100910 (2024).

\bibitem{astrom}
{\AA}str\"om, K.J., Murray, R.M.:
Feedback Systems: An Introduction for Scientists and Engineers. 2nd edn.
Princeton University Press, Princeton, NJ (2021).


\bibitem{bara}
Bara, O., Fliess, M., Join, C., Day, J., Djouadi, S.M.: Toward a model-free feedback control synthesis for treating acute inflammation. J. Theor. Biol. \textbf{448}, 26-37 (2018).

\bibitem{beltran}
Beltran-Carbajal, F., Silva-Navarro, G.:
A fast parametric estimation approach of signals with multiple frequency harmonics. Electr. Power Syst. Res. \textbf{144}, 157-162 (2017).

\bibitem{cartier}
Cartier, P., Perrin, Y.: Integration over finite sets. In: Diener, F.M. (Eds.): Nonstandard Analysis in Practice, pp. 195-204. Springer, Berlin (1995).

\bibitem{coskun}
Coskun, M.Y., Itik, M.:
Intelligent PID control of an industrial electro-hydraulic system. ISA Trans. \textbf{139}, 484-498 (2023).


\bibitem{delvecchio}
Del Vecchio, D., Murray, R.M.: Biomolecular Feedback Systems. MIT Press, Cambridge, MA (2014).

\bibitem{deschutter}
De Schutter, E.: Why are computational neuroscience and systems biology so separate? Plos Comput. Biol. \textbf{4}, e1000078 (2008).

\bibitem{fleming}
Fleming, J.E., Orłowski, J., Lowery, M.M., Chaillet, A.: Self-tuning
deep brain stimulation controller for suppression of beta oscillations:
analytical derivation and numerical validation. Front. Neurosci. \textbf{14}, 639 (2020).

\bibitem{noise}
Fliess, M.: Analyse non standard du bruit. C. R. Acad. Sci. Paris Ser. I \textbf{342}, 797–802 (2006).


\bibitem{mfc1}
Fliess, M., Join, C.: Model-free control. Int. J. Contr. \textbf{86}, 2228-2252 (2013).

\bibitem{mfc2}
Fliess, M., Join, C.: An alternative to proportional-integral and proportional-integral-derivative regulators: Intelligent proportional-derivative regulators.  Int. J. Robust Nonlin. Contr. \textbf{32}, 9512--9524 (2022).

\bibitem{he}
He, D., Wang, H., Tian, Y., Fliess, M.:
MIMO ultra-local model-based adaptive enhanced model-free control using extremum-seeking for coupled mechatronic systems. ISA Trans. \textbf{157}, 233-247 (2025).


\bibitem{chaxel}
Join, C., Chaxel, F., Fliess, M.: ``Intelligent'' controllers on cheap and small programmable devices. Conf. Contr. Fault-Toler. Syst. (SysTol), pp. 554--559. IEEEXplore (2013).
 
\bibitem{join2024epilepsy}
Join C., Jovellar D.B., Delaleau E., Fliess M.: Detection and suppression of epileptiform seizures via model-free control and derivatives in a noisy environment. 12th Inter. Conf.
Syst. Contr., Batna, pp. 79--84, IEEEXplore (2024).

\bibitem{kang}
Kang, J., Huang, X., Xia, C., Huang, D., Wang, F.: Ultralocal model-free adaptive supertwisting nonsingular terminal sliding mode control for magnetic levitation system. IEEE Trans. Indust. Electron., \textbf{71},  5187-5194 (2024).

\bibitem{kuhn2006}
Kühn, A.A., Kupsch, A., Schneider, G.-H., Brown, P.: Reduction in Subthalamic 8-35 Hz Oscillatory Activity Correlates with Clinical Improvement in Parkinson’s Disease. Europ. J. Neurosci. \textbf{23}, 1956–60 (2006).

\bibitem{la}
La Hera, P., Mendoza-Trejo, O., Lideskog, H., Ort\'{\i}z Morales D.: A framework to develop and test a model-free motion control system for a forestry crane. Biomimet. Intell. Robot. \textbf{3}, 100133 (2023).

\bibitem{levenstein}
Levenstein, D, Alvarez, V.A., Amarasingham, A., Azab, H., Chen, Z.S., Gerkin, R.C., Hasenstaub, A., Iyer, R., Jolivet, R.B, Marzen, S., Monaco, J.D., Prinz, A.A., Quraishi, S., Santamaria, F., Shivkumar, S., Singh, M.F., Traub, R., Nadim, F., Rotstein, H.G., Redish, A.D.: On the role of theory and modeling in neuroscience. J. Neurosci. \textbf{43}, 1074--1088 (2023). 

\bibitem{little2013}
Little, S., Pogosyan, A., Neal, S., Zavala, B., Zrinzo, L., Hariz, M., Foltynie, T., Limousin, P., Ashkan, K., FitzGerald, J., Green, A.L., Aziz, T.Z., Brown, P.: Adaptive deep brain stimulation in advanced Parkinson disease. Ann. Neurol. \textbf{74}, 449–457 (2013)

\bibitem{lozano2019}
Lozano, A.M., Lipsman, N., Bergman, H., Brown, P., Chabardes, S., Chang J.W., Matthews K., McIntyre C.C., Schlaepfer T.E., Schulder, M., Temel, Y., Volkmann, J., Krauss, J.K.: Deep brain stimulation: current challenges and future directions. Nat Rev Neurol. \textbf{15},148–160 (2019)

\bibitem{anesth}
Manzoni, E., Rampazzo, M.: Automatic regulation of anesthesia via ultra-local model control, IFAC PapersOnLine {\bf 54-15}, 49--54 (2021).

\bibitem{closedloop}
Mart\'{\i}nez, S., Garc\'{\i}a-Violini, D., Belluscio, M., Piriz, J., S\'anchez-Pe$\tilde{\rm n}$a, R.: Dynamical models in neuroscience from a closed-loop control perspective. IEEE Rev. Biomed. Engin. \textbf{16}, 706--721 (2021). 

\bibitem{mboup}
Mboup, M.: Parameter estimation for signals described by differential equations. Appl. Anal. \textbf{88}, 29–52 (2009).

\bibitem{michel}
Michel, L., Braud, C., Barbot, J.-P., Plestan, F., Peaucelle, D., Boucher, X.: Comparison of different feedback controllers on an airfoil benchmark. Wind Energ. Sci. \textbf{10}, 177–191 (2025).

\bibitem{diab}
MohammadRidha, T., A\"{i}t-Ahmed, M., Chailloux, L., Krempf, M., Guilhem, I., Poirier, J.-Y., Moog, C.-H.: Model free iPID control for glycemia regulation of type-1 diabetes. IEEE Trans. Biomed. Engin. {\bf 65}, 199--206 (2018).


\bibitem{oleary}
O’Leary, T., Sutton, A.C., Marder, E.:
Computational models in the age of large datasets.
Current Opin. Neurobio. \textbf{32}, 87-94 (2015).

\bibitem{nevadoholgado}
Nevado-Holgado, A.J., Terry, J.R., Bogacz, R.: Conditions for the generation of beta oscillations in the subthalamic nucleus–globus pallidus network. J. Neurosci. \textbf{30},12340–-12352 (2010).

\bibitem{othmane}
Othmane, A., Kiltz, L., Rudolph, J.: Survey on algebraic numerical differentiation: historical developments, parametrization, examples, and applications. Int. J. Syst. Sci. \textbf{53}, 1848–1887 (2022).

\bibitem{park}
Park, B., Zhang, Y., Mohammed Olama, M., Kuruganti, T.:
Model-free control for frequency response support in microgrids utilizing wind turbines. Electr. Power Syst. Res. \textbf{194},
107080 (2021).

\bibitem{pasillaslepine}
Pasillas-Lépine, W.: Delay-induced oscillations in Wilson and Cowan’s model: an analysis of the subthalamo-pallidal feedback loop in healthy and parkinsonian subjects. Biol Cybern. \textbf{107}, 289–308 (2013)

\bibitem{pavlides}
Pavlides, A., Hogan, J.S., Bogacz, R.: Improved conditions for the generation of beta oscillations in the subthalamic nucleus–globus pallidus network. Eur. J. Neurosci. \textbf{36}, 2229–2239 (2012).

\bibitem{rosa2015}
Rosa, M., Arlotti, M., Ardolino, G., Cogiamanian, F., Marceglia, S., Di Fonzo, A., Cortese, F., Rampini, P.M., Priori, A.: Adaptive Deep Brain Stimulation in a Freely Moving Parkinsonian Patient. Movement Disorders \textbf{30}, 1003–5. (2015)

\bibitem{rosin2011}
Rosin, B., Slovik, M., Mitelman, R., Rivlin-Etzion, M., Haber, S.N., Israel, Z. Vaadia, E., Bergman, H.: Closed-loop deep brain stimulation is superior in ameliorating parkinsonism. Neuron \textbf{72}, 370-384 (2011).

\bibitem{scherer}
Scherer, P.M., Othmane, A., Rudolph, J.:
Model-free control of a magnetically supported plate.
Contr. Engin. Pract. \textbf{148}, 105950 (2024).

\bibitem{schmidt2024}
Schmidt, S.L., Chowdhury, A.H., Mitchell,  K.T., Peters, J.J., Gao, Q., Lee, H.-J., Genty, K., et al.: At Home Adaptive Dual Target Deep Brain Stimulation in Parkinson’s Disease with Proportional Control. Brain \textbf{147}, 911–22. (2024)


\bibitem{stanslaski2024}
Stanslaski, S., Summers, R.L.S., Tonder, L., Tan, Y., Case, M., Raike, R.S., Morelli, N., Herrington, T.M., Beudel, M., Ostrem, J.L., Little, S., Almeida, L., Ramirez-Zamora, A., Fasano, A., Hassell, T., Mitchell, K.T., Moro, E., Gostkowski, M., Sarangmat, N., Bronte-Stewart, H.: Sensing data and methodology from the Adaptive DBS Algorithm for Personalized Therapy in Parkinson’s Disease (ADAPT-PD) clinical trial. npj Parkinsons Dis. \textbf{10}, 174 (2024).

\bibitem{tagawa}
Tagawa, T., Tamura, T., \"{O}berg, P.A.: Biomedical Sensors and Instruments. CRC
Press, Boca Raton (2011).

\bibitem{ventil}
Truong, C.T., Huynh, K.H., Duong, V.T., Nguyen, H.H., Pham, L.A., Nguyen, T.T.: Model-free volume and pressure cycled control of automatic bag valve mask ventilator. AIMS Bioengin. {\bf 8}, 192--207 (2021).

\bibitem{velisar2019}
Velisar, A., Syrkin-Nikolau, J., Blumenfeld, Z., Trager, M.H., Afzal, M. F. , Prabhakar, V., Bronte-Stewart, H.: Dual threshold neural closed loop deep brain stimulation in Parkinson disease patients. Brain Stimul. \textbf{12}, 868–876 (2019).

\bibitem{yaha}
Yahagi, S., Kajiwara, I.: Data-driven design of model-free control for reference model tracking based on an ultra-local model: Application to vehicle yaw rate control. Proc. Instit. Mechan. Engin., Part D: J. Automob. Engin. \textbf{239}, 1342-1354 (2024).


\end{thebibliography}
\end{document}